\documentstyle[12pt]{article}
\begin{document}
\newcommand{\be}{\begin{equation}}
\newcommand{\ee}{\end{equation}}
\newcommand{\bea}{\begin{eqnarray}}
\newcommand{\eea}{\end{eqnarray}}
\newcommand{\beas}{\begin{eqnarray*}}
\newcommand{\eeas}{\end{eqnarray*}}
\newcommand{\ba}{\begin{array}}
\newcommand{\ea}{\end{array}}
\newtheorem{th}{Theorem}
\newcommand{\tl}{\tilde}
\newcommand{\rar}{\rightarrow}
\newcommand{\lar}{\leftarrow}
\newcommand{\lrar}{\longrightarrow}
\newcommand{\llar}{\longleftarrow}
\newcommand{\fr}{\frac}
\newcommand{\pa}{\partial}
\newcommand{\mb}{\mbox}
\newcommand{\lft}{\lefteqn}
\newcommand{\hs}{\hspace}
\newcommand{\vs}{\vspace}
\newcommand{\hst}{\hspace*}
\newcommand{\vst}{\vspace*}
\newcommand{\lb}{\label}
\newcommand{\nl}{\newline}
\newcommand{\np}{\newpage}
\newcommand{\om}{\omega}
\newcommand{\Om}{\Omega}
\newcommand{\al}{\alpha}
\newcommand{\bt}{\beta}
\newcommand{\eps}{\epsilon}
\newcommand{\veps}{\varepsilon}
\newcommand{\ld}{\lambda}
\newcommand{\Ld}{\Lambda}
\newcommand{\gm}{\gamma}
\newcommand{\Gm}{\Gamma}
\newcommand{\sg}{\sigma}
\newcommand{\bib}{\bibitem}
\newcommand{\ct}{\cite}
\newcommand{\rf}{\ref}
\title{Dimensional Reduction of a Generalized Flux Problem}
\author{\sc Alexander Moroz\thanks{On leave from
 Institute of Physics, Czechoslovak Academy of Sciences,
 Na Slovance 2, CS 180 40 - Prague 8, Czechoslovakia.}\
\thanks{Address after April 1, 1992:
 Institute of Theor. Physics, EPFL,
PHB-Ecublens, CH-1015 Lausanne, Switzerland.}\
\thanks{e-mail address : {\tt moroz@eldp.epfl.ch}}}
\date{\it Institute of Theoretical Physics\\ \it ETH-H\"{o}nggerberg\\
\it CH-8093 Z\"{u}rich, Switzerland}
%
\maketitle
\begin{center}
{\large\sc abstract}
\end{center}
\vspace{0.6cm}
A generalized flux problem with Abelian and non-Abelian fluxes is considered.
In the Abelian case we shall show that the generalized flux problem
for tight-binding models
of noninteracting electrons on either $2n$ or $2n+1$ dimensional
lattice can always be
reduced to a $n$ dimensional hopping problem. A
residual freedom in this reduction enables to identify equivalence
classes of hopping Hamiltonians which have the same spectrum. In the
non-Abelian case the reduction is not possible in general unless the
flux
tensor factorizes into an Abelian one times an element of the
corresponding algebra.
\thispagestyle{empty}
\baselineskip 20pt
\newpage
\noindent
\section{Introduction}
Recently several authors have discussed the generalized flux problem
on a three dimensional lattice \ct{BH,HG,MK,KZ,KH}.
The motivations was either to consider (integer) quantum Hall effect
(QHE) for tight-binding electrons on a Bravais lattice
$\Ld$ in $3$ dimensions ($3D$) with an arbitrary alignement of the magnetic
field $\vec{B}$ \ct{BH,MK,KH}, or an existence and a stability of
flux phases \ct{HG,KZ,HR}.
In \ct{KZ} it has been remarked that in the {\em momentum space} the
generalized flux problem
on a cubic lattice can always be reduced to a one dimensional
hopping problem.
In this letter a similar reduction  is studied
for an arbitrary full rank lattice  $\Ld$ in the $N$-dimensional
{\em direct space}.
To show the {\em dimensional reduction} of a generalized
flux problem we shall generalize one of the building blocks of the
construction given in
\ct{BH,KH}, i.e., that one can always
choose the basis of the Bravais lattice $\Ld$
which contains a given lattice vector
$\vec{b}$ (which of course was assumed to be not a multiple of
any vector of $\Ld$).

The fluxes through oriented elementary faces of a given lattice
form a skew-symmetric tensor $\Phi_{kl}$.
In what follows we shall consider the case when the fluxes
are {\em rational} numbers in the units of $2\pi\Phi_o$, $\Phi_o=hc/e$
being the flux quantum.
Then one part of the problem becomes to be essentially the problem
of defining and then finding {\em normal form for the integral
valued skew-symmetric
matrix} with respect to conjugation by $GL(N,Z)$, the group of
integer valued
matrices $C$ with $|\det C|=1$. Note that a similar
problem for {\em positive definite} matrices $K$ is equivalent
to the problem of a classification of {\em inequivalent} quantum Hall
fluids \ct{NR}. Fortunately, the problem for skew-symmetric
matrices is much more simple and completely solved \ct{Mc}.
By using the {\em normal form of an integer valued skew-symmetric
matrix}  one immediately
finds that a generalized flux
problem on $N$ dimensional lattice, $N$
being either $2n$ or $2n+1$, is characterized
by $n$ different fluxes in general.
Henceforth
we shall call a lattice basis in which the flux tensor
$\Phi_{kl}$ takes its normal form as a $\vec{b}$-basis.
In such a basis $\{\vec{b}_1,\vec{b}_2,\ldots,\vec{b}_N\}$,
the algebra of magnetic translation operators (MTO's)
simplifies since only $n$ pairs of MTO's, $T_{\vec{b}_{2j-1}}$ and
$T_{\vec{b}_{2j}}$, $1\leq j\leq n$, have a non-trivial commutation
relations which of course depend on the canonical fluxes
$\nu_j:=\Phi_{2j-1,2j}$.
Therefore the Hilbert space of states should carry a representation
of $Heis_{\vec{\al}}:=\otimes_{j=1}^n Heis_{\al_j}$,
$Heis_{\al_j}$ being a {\em finite dimensional subgroup} of the Heisenberg
group. Moreover, due to the normal form of a skew-symmetric matrix
with respect to $GL(n,Z)$ conjugation the subgroups $Heis_{\al_j}$
can always be arranged such that if $j<k$ then $Heis_{\al_j}$ is
a subgroup
of $Heis_{\al_k}$.
Thus it is reasonable to take the Hilbert space $W_{\vec{\al}}$ for a $N$
dimensional
$t-b$ model to be the {\em tensor product} of the Hilbert spaces
of {\em theta functions with rational characteristics} $W_{\al_j}$,
$W_{\vec{\al}} =\otimes_{j=1}^n W_{\al_j}$.
The use of theta functions is natural since they
arise in connection with representations of finite dimensional
subgroups of the Heisenberg group \ct{M} which are
relevant for a description of the spectrum of the Bloch electron.
The ordinary theta functions have appeared in the study of a
related problem \ct{FDM}. For the application of {\em theta functions with
characteristics} to the study of the spectrum of
$t-b$ models see \ct{Mo}.
One then shows that in the Hilbert space $W$ the action of a given
$N$-dimensional $t-b$ Hamiltonian ${\cal H}$ of
noninteracting electrons in a magnetic field indeed
reduces to
a $n$-dimensional hopping problem.

We shall show that the dimensional reduction of the generalized flux problem
{\em is not determined uniquely}. For instance the normal form of the
flux tensor is {\em invariant} under the change of the lattice basis
with respect to the action of a {\em transition matrix}
$C=\bar{D}\otimes_s(\oplus_{j=1}^n C_j(\oplus 1))$, $C_j{}'^s$ being
$SL(2,Z)$ matrices and $\bar{D}$ being a translations (see below).
The identity enters the direct sum if and only if $N$ is odd.
The residual symmetry relates different $n$ dimensional hopping
Hamiltonians.
By the construction {\em all the hopping Hamiltonians which belong to the
same orbit of the residual
symmetry group have the same spectrum}.
This just justifies why we are
looking at ``unphysical'' dimensions $D\geq 4$. {\em To find
equivalence classes of $3D$ hopping Hamiltonians one has to look at a
generalized flux problem at $6$ and $7$ dimensions}. Moreover four
dimensional generalized flux problem may be of interest from the
particle physics point of view as zero modes of $2D$ nearest-neighbour
$t-b$ model in
magnetic fields $p/q$, with $q$ being even, are relativistic fermions obeying
the Dirac equation \ct{WZ}, and the model
can be used to provide a lattice regularization of the parity anomaly
\ct{KW}.

The {\em non-Abelian gauge field} of a compact semi-simple Lie group
$G$ poses a more complicated problem:
to define and find a normal form of a skew-symmetric tensor
{\em with values in a corresponding algebra}. Unfortunately even
in the case when the flux tensor takes values in the Cartan subalgebra
a similar reduction, like in the Abelian case, doesn't take place. The
reason is that the problem becomes to be essentially to bring
$r$ different skew matrices {\em simultaneously} to a normal form, $r$
being {\em the rank} of the corresponding algebra.
 The only case when the Abelian-like reduction takes place is
the case a {\em ``uniform flux state''}
which flux tensor  $\Phi_{kl}$ factorizes
according to $\Phi_{kl} =\phi_{kl}\Phi$, where
$\phi_{kl}$ is the (abelian) flux tensor with entries being
rational numbers, and $\Phi$ is an element of the corresponding algebra.

The motivation behind the study of symmetries and spectrum of $t-b$
models in a non-Abelian gauge field is not only purely academic one since
the non-relativistic
quantum mechanic of particles with spin has a natural $U(1)\times
SU(2)$ gauge invariance (see, e.g., \ct{FS}). Other motivation comes
from the mean field approximations to the Heisenberg spin system at
half-filling \ct{Z} having $SU(2)$ gauge invariance \ct{AB}. Finally, an
observation of a general non-Abelian gauge symmetry of the incompressible
Hall fluids at general filling fractions \ct{NR} gives a motivation
to study $t-b$ models in a general non-Abelian
gauge field.
A more detailed discussion
of $t-b$ models we shall give elsewhere \ct{Mor}.

\section{Normal form of a skew symmetric matrix}
\subsection{\bf Abelian case}
\lb{sec:ex}
{\bf 1.} Let us recall some definitions \ct{CS}.
Let $\Ld$ be an arbitrary lattice $\Ld$ of full rank
$N$ with  basis (of primitive translations) $\{\vec{a}_1,\vec{a}_2,
\ldots,\vec{a}_N\}$. It will
be assumed that this is a {\em natural} basis in the sense one that the
integer combinations, $\sum n_j\vec{a}_j$ with $0\leq n_j\leq1$,
of these vectors generate the proximity cell of the lattice $\Ld$.
Let $M$ be a {\em generator matrix} of the lattice $\Ld$, i.e., the matrix
with $j-th$ row being the components of the basis vector $\vec{a}_j$.
Clearly lattice vectors consist of all the vectors $\xi M$ where
$\xi =(\xi_1,\xi_2,\ldots,\xi_n)\in Z^n$. Sets of linearly
independent vectors are related by a regular {\em transition matrix}
$C$, $M'=CM$. $M'$ will be a new generator matrix iff
$\mid\!\det C\!\mid=1$, since both, $C$ and $C^{-1}$, have to have
integer entries as they are supposed to relate two different
lattice bases. Thus $C$ has to be a $GL(N,Z)$ matrix , i.e., modulo an
orientation of a $\vec{b}$-basis
a $SL(N,Z)$ matrix.
We shall say that a matrix $A$
is {\em conjugate} with a matrix $B$ with respect to $GL(N,Z)$ iff exists a
matrix $C\in GL(N,Z)$ such that $A=CBC^T$, $C^T$ being transposed to $C$.
If $\Phi_{kl}$ is some tensor defined on the lattice then a
change of a given basis of $\Ld$ by a transition matrix $C\in GL(N,Z)$
corresponds a conjugation of $\Phi_{kl}$ by $C$.
If the entries of $\Phi_{kl}$ are {\em rational number} it is
always possible to write $\Phi_{kl}$ as $\Phi_{kl}=
\frac{P}{Q}\bar{\Phi}_{kl}=\al\bar{\Phi}_{kl}$, where $Q$ is
{\em the least common
denominator} (lcd) of entries of $\Phi_{kl}$, $P$ is
{\em the greatest common divisor} (gcd) of entries of $Q
\Phi_{kl}$, and $\bar{\Phi}_{kl}$
is an {\em integer valued} matrix. By analogy with the two dimesional
case \ct{Ha}, we have defined the parameter $\al =P/Q \in [0,1)$.
A skew-symmetric
integer valued matrix $\bar{\Phi}_{kl}$ is known to have the following
normal form with respect to the conjugation
by $GL(N,Z)$ :
\begin{equation}
\bar{\Phi} := \left(\ba{cccccc}
\fbox{\rule[-0.2cm]{0cm}{0.6cm}
\(\begin{array}{cc}0&\nu_1\\
-\nu_1&0
\ea\)}
& & & & \\
&  \fbox{\rule[-0.2cm]{0cm}{0.6cm}\(\begin{array}{cc}0&\nu_2\\
-\nu_2&0
\ea\)}
&&&\\
&&\ddots&&\\
&&&\fbox{\rule[-0.2cm]{0cm}{0.6cm}
\(\begin{array}{cc}0&\nu_n\\
-\nu_n&0
\ea\)}
&\\
&&&&(0)
\end{array}\right),
\lb{cform}
\ee
with all remaining entries being zero. The positive integers $\nu_j$
are invariants of $\bar{\Phi}_{\mu\nu}$
under
conjugation. They can be
arranged such that whenever $j\leq k$ then
$\nu_j$ divides
$\nu_k$ (for a constructive proof see \ct{Mc}). Number of nonzero
$\nu_j{}'^s$ is just one-half of
the rank of $\Phi_{kl}$ and $0$ enters the direct sum (\rf{cform})
if only if $N$ is odd.

{\bf 2.} The normal form of a skew-symmetric
matrix (\rf{cform}) doesn't fix $\vec{b}$-basis uniquely.
One finds that the normal form (\rf{cform}) is invariant under
conjugation by
any $SL(N,Z)$ matrix $D$ of the following block form,
\begin{equation}
D:= \left(
\ba{cc}
\ba{cccccc}
\fbox{\rule[-0.2cm]{0cm}{0.6cm} SL(2,Z)}
& & & &\\
& \fbox{\rule[-0.2cm]{0cm}{0.6cm} SL(2,Z)}
&&&\\
&&\ddots&&\\
&&&\fbox{\rule[-0.2cm]{0cm}{0.6cm} SL(2,Z)}
\end{array}
&\ba{c}
d_1\\ d_2\\ \\
\vdots\\ \\
d_{N-1}
\ea
\\
&(1)
\end{array}\right),
\lb{ress}
\ee
with all other entries being zero, and
$\vec{d}:=(d_1,d_2,\ldots,d_{N-1})$
being any integer valued
$(N-1)$-dimensional vector. As before $(1)$ enters $D$ only when $N$ is
odd. By using the Cramer rule one finds that
$D$ matrices form a subgroup
of $GL(N,Z)$ and that any $D$ factorizes into the form
$D= \bar{D}E$, where $\bar{D}$ is nothing but $D$ matrix
{\em with all} $SL(2,Z)$
{\em blocks replaced by identity matrices}, and $E$ is nothing
but $D$ matrix but
{\em with all} $d_j$ {\em set to zero}. One can check that
$\bar{D}$ and $E$
matrices
form subgroups of $D$ (and hence $GL(N,Z)$) and that $D$ is nothing
but a {\em semidirect product} of $E$ and $\bar{D}$,
$D=\bar{D}\otimes_s E$.
\subsection{Non-Abelian case}
 Let us consider the simplest non-Abelian case when the flux tensor
takes values in the corresponding Cartan subalgebra $Cart$ of a
compact semi-simple Lie group of rank $r$.
Let us
take the basis of $Cart$ where all its elements are represented by
real diagonal $r\times r$ matrices. Like in the Abelian case
an element of $Cart$ is said to be {\em rational} iff all its entries are
{\em rational
numbers} in the units of $2\pi\Phi_o$. The flux tensor is then said to be
rational iff all its entries are rational.
The problem of bringing the flux tensor to the form similar to (\ref{cform})
then becomes tantamount to the problem of bringing $r$ different
skew-symmetric
matrices {\em simultaneously} to their normal forms.
This is however {\em impossible} in general unless
all the matrices are proportional each other.
One can argue that it is possible to extend
the
{\em ``mathematical'' residual freedom} (\ref{ress}) of the normal form
of the flux tensor
$\Phi_{kl}$ to a {\em ``physical'' one} since the spectrum of
$t-b$ models is periodic under a change of flux by an integer, i.e.,
under $\alpha\rightarrow\alpha\pm 1$ \ct{Ha}.
Therefore
one can fill up either the upper or the lower triangle block of $D$ by
integers proportional to $Q$ (and still keeping $D$
to be an element of $SL(N,Z)$) {\em without changing the physics}.
Unfortunately neither the ``mathematical'' nor ``physical'' residual freedom
of the normal form (\ref{cform}) help. The reason is that the
action of an $SL(N,Z)$ matrix on an integer valued vector $\vec{A}\in Z^N$
{\em preserves
the greatest common divisor of its components}, which is the (only)
invariant of $\vec{A}$ under the action of $SL(N,Z)$.
Therefore the conjugation
by $SL(N,Z)$ matrices preserves the greatest common divisors of rows
and columns of a given matrix. Thus (\ref{cform}) is not a normal form
(if any exists at all) of  the Cartan subalgebra valued flux tensor.
The only case when the general flux tensor has its normal form
like (\ref{cform}) is the case of a {\em uniform flux state}, where
$\Phi_{kl}=\phi_{kl}\Phi$, $\phi_{kl}$ being an Abelian
flux tensor and $\Phi$ an element of the Lie algebra. Note that the
nonexistence
of the Abelian like normal form (\ref{cform}) even for the Cartan subalgebra
valued flux tensor doesn't
preclude us to consider the case when the flux tensor has already
the Abelian like normal form (\ref{cform}) (with $\nu_j{}'^s$ replaced
by general fluxes) in the normal coordinates of $\Ld$.

\section{Symmetries and spectrum of $t-b$ models}
Let us take the affine coordinates $(x_1,x_2,\ldots,x_N)$ in $R^N$ along the
vectors of the
natural lattice basis. The scalar product (metric) in this
(non-cartesian) coordinates
is given by the {\em Gram matrix} $G$ of the lattice $\Ld$, $G=MM^{tr}$:
$\vec{x}.\vec{y}=G_{kl}x^k y^l$, and the derivatives $\partial_j{}'^s$
satisfy the usual commutation relation, $[\partial_j,\partial_k]=0$.

Let ${\cal D}_j$ be the {\em covariant derivatives} (a multiple of the
standart  {\em velocity operator}),
${\cal D}_j =\partial_j - 2\pi iA_j/\Phi_o$.
Let us  assume that the
flux tensor, $\Phi_{kl}$, of a given gauge field
doesn't depend on coordinates. For a given vector
$\vec{a}=\sum_j n_j\vec{a}_j$ it is always possible to take the corresponding
shift operator $S_{\pm \vec{a}}$ as
$S_{\pm\vec{a}} = e^{\pm \sum_j n_j{\cal D}_j}$.
Since in our coordinates the commutator of covariant derivatives is
proportional to the flux $\Phi_{kl}$ through an oriented face $(kl)$,
$[{\cal D}_k,{\cal D}_l]=- 2\pi i \Phi_{kl}$, the shift operators
satisfy the algebra
\be
S_k S_l =S_l S_k e^{-2\pi i \Phi_{kl}}.
\lb{sho}
\ee
This is most easily seen by taking as a  ``basis'' of shift operators
the shift operators along the
vectors of a $\vec{b}$-basis  in which the flux tensor
$\Phi_{jk}$  takes its normal form and it is nonzero only for pairs of
indices $(2k-1,2k)$, $1\leq k\leq n$.
Let $(\bar{x}_1, \bar{x}_2,\ldots,\bar{x}_N)$  be affine
coordinates  along vectors of a given $\vec{b}$-basis.
Then one can take
$\bar{\cal D}_{2k-1} =\bar{\partial}_{2k-1} - 2\pi i\al_k
\bar{x}_{2k}$, and $\bar{\cal D}_{2k} =\bar{\partial}_{2k}$,
where
the parameter $\al_k$ is defined according to $\al_k
:=P\nu_k/Q:=p_k/q_k$, $p_k$ and $q_k$ being relative prime integers.
The shift operators provides projective representation of
the group of lattice translations with $2$-{\em cocycle}
$\om_{kl}= -\pi \Phi_{kl}$ and the basis shift operators in the normal
basis can
be expressed through the basis shift operators in $\vec{b}$-basis and
vice versa.

A general $t-b$ model is then defined by the Hamiltonian
\begin{equation}
{\cal H} := \sum_{\vec{a}} t_{\vec{a}}S_{\pm \vec{a}} + h.c.
\label{Ham}
\end{equation}
The (finite) sum is over the vectors $\vec{a}\in\Ld$ with
$\|\vec{a}\|\leq \eta$, $\eta$ being some real number, and
$t_{\vec{a}}=t_{-\vec{a}}$.
Clearly, this also includes $t-b$ models
with direction dependendent overlap, $t_{\vec{a}}\neq t_{\vec{a}'}$,
in general.
The $t-b$ Hamiltonians (\ref{Ham}) are {\em single band effective
Hamiltonians} for the Schr\"{o}dinger Hamiltonian $H$,
$H=(\vec{p}-(e/c)\vec{A})^2 +V(\vec{r})$, $V(\vec{r})$ being a
periodic potential where the ``effectiveness'' depends on the Wannier
function of the single band in zero magnetic field \ct{N}.

Since the flux tensor is constant there exists a gauge
transformation $G_{\vec{a}}$ such that ${\cal H}(\vec{r}+\vec{a})$ is gauge
equivalent to ${\cal H}(\vec{r})$,
${\cal H}(\vec{r}+\vec{a})=G^{-1}_{\vec{a}}{\cal H}(\vec{r})G_{\vec{a}}$.
The operators $T_{\vec{a}}:= G_{\vec{a}}
e^{\partial_{\vec{a}}}= G_{\vec{a}}T^o_{\vec{a}}$, called operators of
magnetic translations (MTO's), commute
with ${\cal H}$ \ct{B}, $T^o_{\vec{a}}$ being the
usual translation (in zero magnetic field).
One finds that in a $\vec{b}$-basis the MTO's can be written as exponentials
of {\em covariant derivatives} with some {\em fake gauge
field} $\vec{\tl{A}}$,
$T_{\vec{a}_j}=\exp{\tl{\cal D}}_j$,
$\tl{{\cal D}}_j=\partial_j - 2\pi i\tl{A}_j/\Phi_o$.
We shall refer to such MTO's as {\em the lattice MTO's}.
Again it is seen more easily in a $\vec{b}$-basis where
$\tl{\cal D}_{2k-1}=\bar{\partial}_{2k-1}$, and
$\tl{\cal D}_{2k}=\bar{\partial}_{2k} - 2\pi i\al_k\bar{x}_{2k-1}$,
$1\leq k\leq n$.
One can check that there doesn't exist a gauge transformation
which relates the shift operators with MTO's. Thus MTO's are independent of
the shift operators. The
components of the fake gauge potential $\vec{\tl{A}}$ are
such that $\forall\ j,k$ : $[\bar{\cal D}_j,\tl{\cal D}_k]=0$.
The MTO's satisfy the algebra where the only nontrivial commutators
are
\be
T_{\vec{b}_{2k-1}}T_{\vec{b}_{2k}} = T_{\vec{b}_{2k}}
T_{\vec{b}_{2k-1}} e^{2\pi i\al_k} ,
\hspace{2cm} T_{\vec{b}_{2k-1}}T_{\vec{b}_{2k}} =
T_{\vec{b}_{2k-1}+\vec{b}_{2k}} e^{i\pi\al_k} .
\lb{c-mt}
\ee
They provide a projective representation of the group
of lattice translations with the $2$-cocycle
$\tl\om_{kl}=-\om_{kl}=\pi\Phi_{kl}$ \ct{B}.
{}From now on the whole buisness is almost the same as for $2D$
models. One finds that any pair (\ref{c-mt}) of MTO's provides
an irreducible representation of a finite Heisenberg group
$Heis_{\al_k}$ \ct{Mo,B} and the whole Hilbert space has to
carry a reperesentation of $Heis_{\vec{\al}}=\otimes^n_{k=1}Heis_{\al_k}$.
One picks up the maximal set of commuting operators
$\{T_{\vec{b}_{2k-1}},
T^{q_k}_{\vec{b}_{2k}}|1\leq k\leq n\}$, and defines a magnetic
Bravais lattice $\Ld^m=\oplus\ld^m(\oplus\vec{b}_N$ if $N$ is odd)
where any $\ld^m$ has the basis $\{\vec{b}_{2k-1},
q_k\vec{b}_{2k} \}$, $1\leq k\leq n$.
A single band will split into $\det\Ld^m/\det\Ld=\prod_k q_k$
subbands and the corresponding subbands are $\prod_k q_k$-fold
degenerated. The above conclusions hold for either the Schr\"{o}dinger
Hamiltonian $H$ or for a $t-b$ Hamiltonian ${\cal H}$. However there
are some differences between the two Hamiltonians.
Note that the algebra (\ref{c-mt}) implies that the
``minimal'' commuting
MTO's  are
$T_{\vec{b}_{2k-1}}$ and $T_{q_k\vec{b}_{2k}/p_k}$. However,
any $T_{q_k\vec{b}_{2k}/p_k}$ is not the lattice translations and
doesn't commute with $H$. On the other hand they commute with
${\cal H}$ but they
don't relate points which are related by ${\cal H}$ and hence
is unimportant also for the classification of the spectrum of ${\cal
H}$, although from a different reason.
By a simple consideration one finds that $1/\al$ roots (powers) of
the shift operators defined in a $\vec{b}$-basis, i.e.,
{\em the Azbel operators} $\hat{A_j}{}'^s$  \ct{Ha},
$\hat{A_{2k-1}} := S_{2k-1}^{1/\al_k}$, $\hat{A_{2k}} := S_{2k}^{1/\al_k}$
also commute with the Hamiltonian ${\cal H}$. From the similar reasons
as above they are not relevant for the spectrum of either $H$ or
${\cal H}$. All these operators reflects the periodicity
of the Brillouin zone. The most important difference between
$H$ and ${\cal H}$ is that $H$ commutes with the lattice MTO's only
while ${\cal H}$ commutes with the continuous MTO's,
$T_{\xi\vec{b}_k}=\exp{\xi\tl{\cal D}}_k$, $\forall \xi\in R$,
 as well, as
a consequence of $[\bar{\cal D}_j,\tl{\cal D}_k]=0$ \ct{Mo}.
Therefore we shall look for an appropriate Hilbert space, the lattice
MTO's has to be always considered as a subgroup of the full  Heisenberg
group \ct{M,Mo}.
These additional symmetries of ${\cal H}$ with regard to $H$ don't
lead to any additional degeneracy of energy levels since
they don't
relate points which are related by the Hamiltonian ${\cal H}$.
They result a {\em continuous family
of equivalent Hilbert spaces parametrized by points of unit cell of $\Ld$}.
They differ from each other by the boundary
conditions imposed (see below; for a similar case see \ct{FH}).
The other way around is to say the spectra of $H$ and ${\cal H}$
concides as sets but have different multiplicities,
$H$ being a {\em direct integral} of ${\cal H}$ over the equivalent
Hilbert spaces \ct{N}.

\section{Hilbert space}

As it has been said above the Hilbert space should carry a
representation $Heis_{\vec{\al}}=\otimes^n_{k=1}Heis_{\al_k}$, and
that
this group has to be considered as a subgroup of a  tensor
products of Heisenberg groups $Heis$. We
shall recall that the
he Heisenberg group $Heis$, which is nothing but the central extension
of the group of ordinary lattice translations, can be defined as a
set of elements
$\{(\lambda,a,b) | \lambda\in C^*; a,b \in R\}$ with the
multiplication law:
$(\lambda ,a,b)(\lambda ', a',b')=(\lambda\lambda '
e^{-2\pi i (ba'-b'a)},a+a',b+b'),$
where $C^*$ is a unit circle in the complex plane. Let us denote
by $q\Gamma$ a discrete subgroup of $Heis$,
$q\Gamma :=\{(1,qa,b) | a,b \in Z \}$.
We shall denote by $V$ the Hilbert space of entire functions
$f(z)$ with the norm induced by the scalar product,
\be
||f||^2 = \int \exp (-2\pi y^2 /Im\tau )|f(x+iy)|^2\  dx\ dy,
\lb{norm}
\ee
where the integral is taken over the elementary periodicity domain,
$\tau$ being a modular parameter (modulus).
Let $V_q$ be a  subspace of $V$
invariant under $q\Gamma$. The action of
$(1,qa,b)\in q\Gamma$ on $f(z)\in V_q$ is given as usual,
$(1,qa,b)f(z)= e^{\pi ib^2\tau +2\pi ib(z+qa)}f(z+qa+b\tau),$
and $f(z)\in V_q$ if and only if
$f(z) = \sum _{n\in (1/q)Z} c_n \nl\exp \{\pi in^2\tau + 2\pi inz\},$
with $c_{n+1} =c_n$, i.e., $V_q$ is a $q-$dimensional (complex)
subspace of $V$.
Then a discrete subgroup $Heis_q$,
$Heis_q := \{(\lambda,a,b) |\lambda\in C^*_q, a\in Z, b\in (1/q)Z\}
/$(mod $q\Gamma$) = $C^*_q\times Z/qZ \times (1/q)Z/Z $,
$C^*_q$ being the cyclic group of $q-$roots of 1, commutes with
$q\Gamma$. Following readily arguments in \cite{M} step by step
one can show that the finite
group $Heis_q$ acts irreducibly on $V_q$. Moreover, one has even
an analoque of the Stone-von Neumann theorem for discrete subgroups
of the Heisenberg group \cite{M} .
Because of irreducibility the action of $Heis_q$ on $V_q$ determines a
canonical basis for $V_q$ and $Heis_q$ acts in a fixed way. The standart
basis of $V_q$ is given in terms of theta functions with
upper characteristic $\ell\alpha$ (modulo a constant), where
$\ell = 0,1,\ldots ,q-1$ \ct{M}.
For a given modulus $\tau_k$ and for a given pair
$(\bar{x}_{2k-1},\bar{x}_{2k})$ of coordinates let the complex
variable $z_k$ be defined by $z_k= z_k (\tau_k):= \tau_k\al_k \bar{x}_{2k}$ and
let us consider $\al_k\bar{x}_{2k-1}$ as the lower characteristic of the
modified theta function
\( g[\stackrel{\ell\al}{\scriptstyle x/a_1}]
(\tau |\tau_k\al_k \bar{x}_{2k}) \)
defined for any $Heis_{\al_k}$ as follows,
\be
g [\stackrel{\ell\al_k}{\scriptstyle \al_k\bar{x}_{2k-1}}]
(\tau_k |\tau_k\al_k ) :=
e^{\pi i\tau_k \bar{x}_{2k}^2\al_k^2}\theta
[\stackrel{\ell\al_k}{\scriptstyle \al_k\bar{x}_{2k-1}}](\tau_k |z_k(\tau_k)).
\lb{bf}
\ee
$\theta [\stackrel{\ell\al_k}{\scriptstyle \al_k\bar{x}_{2k-1}}]
(\tau_k |\tau_k\al_k\ \bar{x}_{2k})\in V_{\al_k}$
is the usual Jacobi theta function with characteristics
\ct{M},
\begin{equation}
\theta [\stackrel{\ell\al}{\scriptstyle x}]
(\tau |\tau\al y) = \sum_{n=-\infty}^{\infty}
\exp\{\pi i\tau (n+\ell\alpha )^2 + 2\pi i(n+\ell\alpha )
(x+\tau \alpha y)\}.
\label{theta}
\end{equation}
With slight abuse of notation we shall write $(\tau,\al,\vec{b}_1,
\vec{b}_2,x,y)$ instead of
$(\tau_k,\al_k,\nl\vec{b}_{2k-1},\vec{b}_{2k},
\bar{x}_{2k-1},\bar{x}_{2k})$ whenever doesn't threat a
confusion.
In the following step we shall introduce the momenta $(k_1,k_2)$ and
define functions
\be
g^{k_1k_2}
[\stackrel{\ell\al}{\scriptstyle x}]
(\tau|\tau\al y) := e^{ik_1x +ik_2y}g[\stackrel{\ell\al}{\scriptstyle x}]
(\tau|\tau\al y)
\lb{g-fun}
\ee
Let us denote the Hilbert space of functions
$g^{k_1k_2} [\stackrel{\ell\al}{\scriptstyle x}]$
with the scalar product (\ref{norm})
by $W_\al$.
For a given modulus $\tau$ the lattice MTO's (considered
as a subgroup of
the Heisenberg group) provide the
unitary irreducible
representation (UIR) of $Heis_q$ in $W_\al$,
$Heis_q \ni (1,\pm 1,0)
\hookrightarrow T_{\pm\vec{b}_1}$,
$Heis_q\ni (1,0,\pm\alpha )
\hookrightarrow T_{\pm\vec{b}_2}$.
An interesting point is that if one takes the basis functions
(\ref{theta}) with $(x,y)\rightarrow (y,x)$, $(y,-x)$, or $(-y,x)$,
then (\ref{g-fun}) still provides UIR (\ref{c-mt}) of MTO's.
Moreover one can take as a basis of $Heis_{\al_k}$ the functions
$g^{k_1k_2} [\stackrel{\ell\al}{\scriptstyle x}]$ transformed by a
{\em modular transformation}. The transformations
are of the form,
\be
\tau\hookrightarrow \dot{\tau} = \frac{d\tau +b}{c\tau +a},\hs{1cm}
z\hookrightarrow \dot{z} =
\frac{z}{c\tau +a} ,
\lb{mod}
\ee
the parameters $a,b,c,d$ being integers, $ad-bc=1$,
and they form {\em the group of modular transformations}, $PSL(2,Z)$.
Under this transformations functions (\ref{g-fun}) transforms as
follows \ct{Mo},
\be
g^{k_1k_2}[\stackrel{\ell\al}{\scriptstyle x+b\al y}](\tau |\tau\al ay) =
(c\tau +d)^{-1/2}u_\ell e^{-i\pi ab\al^2 y^2}
g^{k_1k_2}[\stackrel{\dot{\ell}\al}{\scriptstyle \dot{x}}]
(\dot{\tau}|\dot{\tau}\al y),
\lb{modtr}
\ee
where $\dot{\ell}\al =\ell\al d -cx +cd/2$, $\dot{x}=
ax -b\ell\al +ab/2$, and $u_\ell$ is a phase factor which doesn't depend
on $\tau$ and $y$.

Note that in a rational magnetic field, $\al =p/q$, irreducible action
of the lattice MTO's  requires that the Hilbert
space be specified by periodic boundary conditions (PBC) on the square
defined by translations by $q$ {\em lattice spacings in both directions}
despite that already $T_{\vec{a_1}}$ and
$T_{q\vec{a_2}}$ commute (\rf{c-mt}).
Let us look at the behaviour of $g[\stackrel{\ell\al}
{\scriptstyle x}](\tau |\tau\al y)$ under lattice
translations,
\begin{eqnarray}
g[\stackrel{\ell\al}{\scriptstyle x\pm 1}]
(\tau |\tau\al y) &=& e^{\pm 2\pi i\ell\al}g[\stackrel{\ell\al}{\scriptstyle
x}]

(\tau |\tau\al y) ,
\nonumber \\
g[\stackrel{\ell\al}{\scriptstyle x}]
(\tau |\tau\al (y\pm 1))
&= & e^{\mp i2\pi\al x}
g[\stackrel{\ell\al}{\scriptstyle x}]
(\tau |\tau\al y).
\lb{xpc}
\eea
Thus basis functions
$g^{k_1k_2}[\stackrel{\ell\al}{\scriptstyle x}]
(\tau |\tau\al y)$ satisfy {\em the same boundary conditions} (BC) on the
elementary periodicity $q\times q$ square at any point of
a given  $(x_o +m,y_o +n)$ lattice.
This corresponds to the fact that
these points are connected by the Hamiltonian ${\cal H}$ and states
on this lattice belong to {\em the same} Hilbert space.
Because of (\ref{xpc})
any basis function $g[\stackrel{\ell\al}{\scriptstyle x}]
(\tau |\tau\al y)$
carries  {\em internal Bloch momenta}
$k_1=2\pi\ell\al$ and
$k_2=-2\pi p x_o$ on a given $\ld^m_{(x_o,y_o)}$.
In  $2D$ case one takes a  combination
$\psi (x,y)=
e^{ik_1x +ik_2y}u(x,y)$, where $u (x,y):=
\sum_\ell d_\ell e^{-2\pi i\ell\al x  +2\pi i\al x_o y}
g[\stackrel{\ell\al}
{\scriptstyle x}](\tau |\tau\al y)$ and let ${\cal H}$ act on it.
The action
of ${\cal H}$ leads on the Harper equation (HE) \ct{Ha} for
the components $d_\ell$ \ct{Mo}. One then diagonalizes
$T_{\vec{b}_1}$ in the corresponding degenerate subspace of $W_\al$ and
finds eigenstates $\psi (x,y)$ of ${\cal H}$ which have definite
Bloch momenta $(k_1,k_2)$ on  $\ld^m$.
{\em All the symmetries of the spectrum of} $t-b$
{\em models} \ct{WZ} {\em then appear
very naturally as a consequence of the periodicity properties}
(\ref{xpc})
{\em of the modified theta functions} (\ref{g-fun}) \ct{Mo,Mor}.

In the general case one takes
$\psi (x,y)= \prod_{s=1}^n
e^{ik^s_1x +ik^s_2y}u_s(x_s,y_s)$, where $u_s (x_s,y_s):=
\sum_\ell d_\ell e^{-2\pi i\ell\al_s x_s  +2\pi i\al x^o_s y_s}
g[\stackrel{\ell\al_s}
{\scriptstyle x_s}](\tau_s |\tau_s\al_s y_s)$, and index $s$ refers
to a corresponding subspace $Heis_{\al_s}$.
To show the dimensional reduction let us consider the action of
a basis shift operator $S_{\vec{a}}$ in the natural basis on a basis
function $g[\stackrel{\ell\al}
{\scriptstyle x}](\tau |\tau\al y)$ (\ref{g-fun}), expressed in
the $\vec{b}$-basis. If $C$ is a transition matrix from the natural
to a $\vec{b}$-basis,
$\vec{a}_k=C^{-1}_{kl}\vec{b}_l$, $\bar{x}_k= x_l C^{-1}_{lk}$,
and $\bar{\partial}_k= C_{kl}\partial_l$. Therefore
\be
S_{\vec{a_l}} \prod_s g[\stackrel{\ell\al_s}
{\scriptstyle x_s}](\tau_s |\tau_s\al_s y_s)\}=
\prod_s g[\stackrel{\ell\al_s}
{\scriptstyle x_s+ C^{-1}_{l,2s-1}}](\tau_s |
\tau_s\al_s (y_s+ C^{-1}_{l,2s})).
\lb{red}
\ee
Now, by using  the periodicity properties (\ref{xpc}) of $g[\stackrel{\ell\al}
{\scriptstyle x}](\tau |\tau\al y)$ under lattice translations one
gets  $n$ dimensional hopping
Hamiltonian. Since we have a residual freedom given by the group $D$,
one can change $C^{-1}_{kl}$ in (\ref{red}) to $(DC)^{-1}_{kl}$. This
results
in a seemingly different hopping Hamiltonian. By the construction it
has to have the same spectrum as the previous one. Thus by starting from
a given flux problem one can obtain the whole family
({\em equivalence class})
of equivalent hopping
Hamilton
ians. Similarly, by taking basis functions transformed
under a modular transformation one obtains another hopping
Hamiltonians.
We shall follow this interesting observation in more
details elsewhere \ct{Mor}.

\section{Conclusion}
We have just shown that a generalized flux problem with Abelian
fluxes of tigt-binding models
of noninteracting electrons on either $2n$ or $2n+1$ dimensional
lattice can always be
reduced to  $n$ dimensional hopping problem. We have observed that
the Hilbert space ``factorizes'' into the tensor product
of Hilbert spaces for $2$ (or $3$) dimensional flux problems. This
factorization enables rather easy to discuss the group theoretical
properties of the spectrum  of a given flux problem. The
properties of the spectra
holds for the (more fundamental) Schr\"{o}dinger Hamiltonian $H$
as well.
The reason that
we were looking at ``unphysical'' dimensions $D\geq 4$ is justified
by the observation that there is some
residual freedom in this reduction. This freedom  enables to identify
equivalence
classes of hopping Hamiltonians which have the same spectrum in the
``physically'' interesting dimensions.
The group of residual freedom contains $SL(2,Z)$ as a subgroup.
In the
non-Abelian case the reduction is not possible in general unless the
flux
tensor factorizes into an Abelian one times an element of the
corresponding algebra.

In the present consideration we have also encountered the {\em modular
symmetry} $PSL(2,Z)$ as we have the freedom
to take modular parameter of the basis functions (\ref{g-fun}) of
a corresponding Hilbert subspace at our own choice.
We should like to understand what role the symmetry does play
for the $t-b$ models. We should like to mention that
similar symmetry, although different in origin, has appeared in a
similar problem considered by Wen \ct{W},
where the modular invariance $PSL(2,Z)$
relates models with {\em different effective mass
tensor in the coupling
constant space}. An invariance under some subgroup of
$PSL(2,Z)$  has been also observed
recently in an dissipative extension of the
Hofstadter model \ct{CF}.
A detailed consideration of the problem we shall give elsewhere
\ct{Mor}.\vs{0.4cm}\\

{\bf Acknowledgements}\vs{0.6cm}\\
I should like to thank T. Kerler and E. Thiran for many fruitful
discussions.
Financial support by the Swiss National Foundation
is also gratefully acknowledged.\vs{0.6cm}\nl
{\bf Note added}: As it was told to me by
A. P. Polychronakos the algebra like (\rf{c-mt})
is also satisfied by the physical observables of the pure
Chern-Simons theory on the torus for a collection of compact abelian
gauge fields \ct{Po}. I thank him for this remark.
\np

\end{document}